\pdfoutput=1
\documentclass[apj]{emulateapj}
\newcommand{\beq}{\begin{equation}}
\newcommand{\eeq}{\end{equation}}

\newcommand{\citei}[1]{\citeauthor{#1} \citeyear{#1}}

\newcommand{\hi}{H{\sc i}~}
\newcommand{\hia}{H{\sc i}}
\newcommand{\kms}{km ${\rm s^{-1}}$~}
\newcommand{\kmsa}{km ${\rm s^{-1}}$}
\usepackage{rotating}
\usepackage{footnote}
\slugcomment{Accepted to the ApJS}
\begin{document}
\title{The GALFA-H\hspace{0.15em}{\sc i} Survey: Data Release 1}
\author{J.~E.~G.~Peek\altaffilmark{1}, Carl Heiles\altaffilmark{2}, Kevin A. Douglas\altaffilmark{3}, Min-Young Lee\altaffilmark{4}, Jana Grcevich\altaffilmark{1}, Sne\v zana Stanimirovi\'c\altaffilmark{4}, M.~E.~ Putman\altaffilmark{1}, Eric J. Korpela\altaffilmark{5}, Steven J. Gibson\altaffilmark{6}, Ayesha Begum\altaffilmark{4}, Destry Saul\altaffilmark{1}, Timothy Robishaw\altaffilmark{7}, Marko Kr\v co\altaffilmark{8}}


\altaffiltext{1}{Department of Astronomy, Columbia University, New York, NY 10027, USA}
\altaffiltext{2}{Radio Astronomy Lab, UC Berkeley, 601 Campbell Hall, Berkeley, CA 94720}
\altaffiltext{3}{School of Physics, University of Exeter, Stocker Road, Exeter, United Kingdom EX4 4QL}
\altaffiltext{4}{University of Wisconsin, Madison, 475 N Charter St, Madison, WI 53703}
\altaffiltext{5}{Space Sciences Laboratory, University of California, Berkeley, CA 94720}
\altaffiltext{6}{Department of Physics and Astronomy, Western Kentucky University, Bowling Green, KY 42101}
\altaffiltext{7}{Sydney Institute for Astronomy, School of Physics, The University of Sydney, NSW 2006, Australia}
\altaffiltext{8}{Department of Astronomy, Cornell University, Ithaca, NY 14853}
\begin{abstract}
We present the Galactic Arecibo L-Band Feed Array \hi (GALFA-\hia) survey, and its first full data release (DR1). GALFA-\hi is a high resolution ($\sim$4$^\prime$), large area (13000 deg$^{2}$), high spectral resolution (0.18 \kmsa), wide band ($-700 < v_{\rm LSR} < +700$ \kmsa) survey of the Galactic interstellar medium in the 21-cm line hyperfine transition of neutral hydrogen conducted at Arecibo Observatory. Typical noise levels are 80 mK RMS in an integrated 1 \kms channel. GALFA-\hi is a dramatic step forward in high-resolution, large-area Galactic \hi surveys, and we compare GALFA-\hi to past, present, and future Galactic \hi surveys. We describe in detail new techniques we have developed to reduce these data in the presence of fixed pattern noise, gain variation, and inconsistent beam shapes, and we show how we have largely mitigated these effects. We present our first full data release, covering 7520 square degrees of sky and representing 3046 hours of integration time, and discuss the details of these data.
\end{abstract}

\keywords{ISM:\hia; atlases; surveys; radio emission lines}

\section{Introduction}
In 2004 the Arecibo L-Band Feed Array (ALFA) was installed on the Arecibo 305-meter radio telescope. While Arecibo's status as the largest single-dish antenna in the world had always made it a premier instrument in cm-wave radio astronomy, its ability to map large areas of sky had been hindered by the fact that it could only observe one position at a time. One can consider the original L-band wide receiver, which covers a similar frequency range to ALFA, as a `single-pixel camera' with a single $\sim3.5^\prime$ beam. ALFA, by comparison, monitors 7 positions at once, thus increasing the mapping speed of Arecibo by a factor of 7. This tremendous boost in efficiency led many different science teams to consider for the first time the possibility of mapping huge swaths of sky with Arecibo to create a new generation of high-fidelity, high-resolution sky atlases. These groups were divided by topic: pulsar (P-ALFA; \citei{Cordes08}), extra-galactic (E-ALFA; \citei{Giovanelli05}, \citei{Henning10}, \citei{Auld06}, \citei{Freudling08}), extraterrestrial intelligence \citep{Korpela09}, and Galactic science (GALFA). The GALFA consortium itself consists of three independent groups, or sub-consortia, divided by the radiative processes they measure: radio recombination lines (GALFA-RRL; \citealt{Terzian05}), continuum radiation (GALFACTS; \citei{GT09}) and the 21 cm line transition of \hi (GALFA-\hia; \citealt{Goldsmith03}, this work). 

Successful large \hi surveys of the Galaxy have been undertaken in the past, perhaps beginning in the modern era with the pioneering work of \citet{WW73} and \citet{HH74}. More recently there has been an explosion of Galactic \hi surveys. The Leiden-Argentine-Bonn (LAB) Survey \citep{Kalberla05} is a full sky, stray-radiation corrected map of the Galaxy in \hi at relatively low angular resolution (36$^\prime$). The International Galactic Plane Survey (IGPS; \citealt{Taylor03}, \citealt{Stil06}, and \citealt{McC-G05}) is a compilation of relatively high resolution ($\sim1 ^\prime$) interferometric observations of the Galactic plane. The Galactic All Sky Survey (GASS; \citealt{McC-G09}), is a half-sky intermediate resolution (15$^\prime$) survey in the southern hemisphere. GALFA-\hi combines the large-area approach of GASS and LAB with resolution only somewhat coarser than the IGPS. These qualities, along with high spectral resolution and superb fidelity, allows GALFA-\hi to contribute to a whole new range of science in and around our Galaxy. A number of other surveys planned or in progress are also tackling the imaging of large areas of the Galactic sky in \hia. We have tabulated the main parameters of these completed, ongoing, and planned surveys in Table \ref{survcomp} for ease of comparison.

The GALFA-\hi survey was designed to study a broad range of topics under the umbrella of the neutral hydrogen, Galactic interstellar medium (ISM). To date, many different scientific endeavors have been undertaken with GALFA-\hia. We have studied gas in the Galactic halo, with projects based on observations of high-velocity clouds (\citei{Peek07}, \citei{Stanimirovic08}, \citei{Peek09}, and \citei{Hsu10}) as well as constraints on the scale of gas outflows from stars in globular clusters \citep{VanLoon09}. We have also investigated gas much closer to home, with studies of turbulence in the local ISM \citep{Chepurnov10} and studies of cold neutral medium clouds close to the sun \citep{PHMP10}. Compact clouds, small \hi clouds distinct from diffuse background emission, are a focus of investigation (\citei{Begum10}, \citei{Saul10}). GALFA-\hi has also been leveraged to study the complexities of the Galactic plane in \hi (\citei{Krco07}, \citei{Gibson10}), as well as the details of galaxies in the Local Group (\citealt{Putman09}, \citealt{Grcevich08}). Excluding the Galactic plane work, all of these projects are based on data that is presented to the public in this work.

\begin{sidewaysfigure}
\begin{center}
\includegraphics[scale=.25, angle=0]{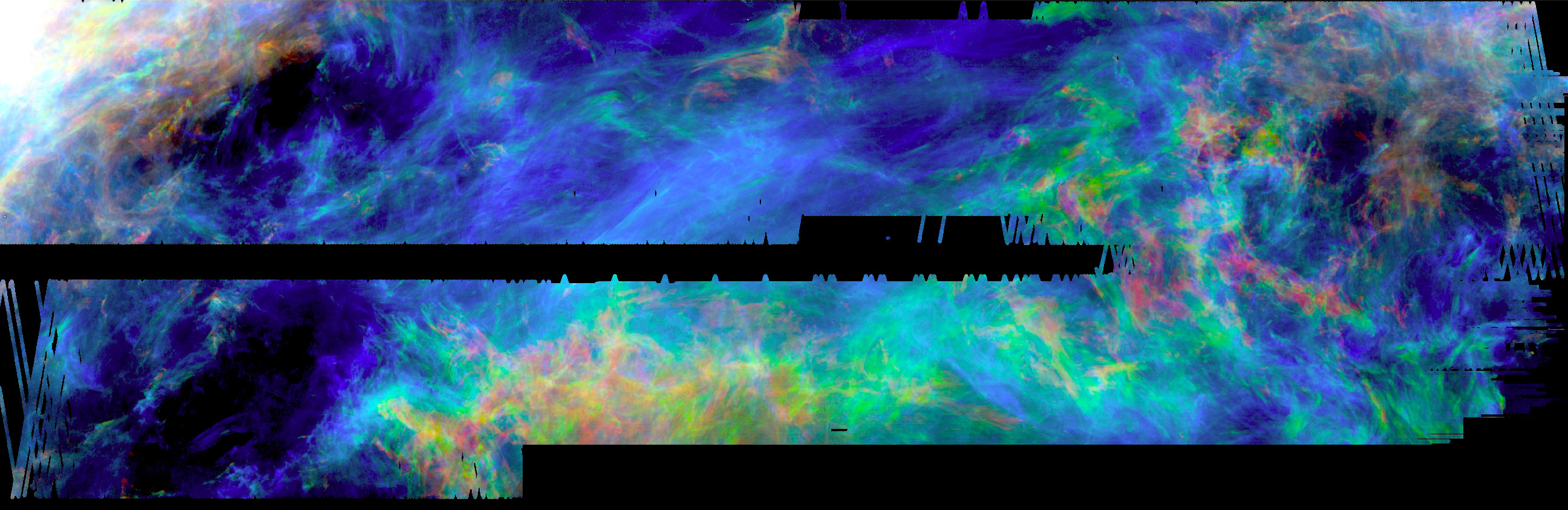}
\caption{A section of the DR1 data, 120$^\circ$ in $\alpha$ by 39$^\circ$ in $\delta$, towards $\left( \alpha, \delta \right) = \left(20^\circ, 18.6^\circ \right)$ . The red, green, and blue channels represent -12.8, -9.2 and -5.5 \kmsa, respectively. \label{pretty}}
\end{center}
\end{sidewaysfigure}

In this paper we present a description of the GALFA-\hi survey in progress and the first data release (DR1).  DR1 covers 7520 square degrees of sky, utilizing 3046 hours of Arecibo data taken with ALFA. A large fraction of this data was taken commensally with other projects, allowing for the efficient use of telescope time. The survey covers the velocity range $-700 < v_{\rm LSR} < +700$ \kms at 0.184 \kms spectral resolution. Due to the differing integration times over the survey footprint (see Figure \ref{coverage}), RMS noise in a 1 \kms channel ranges from 140 mK to 60 mK, with a median value of 80 mK. The final effective beamsize is 3.9 arcminutes in RA and 4.1 arcminutes in Declination, slightly larger than the raw ALFA beamsize. 
Figure \ref{pretty} shows an area covering roughly 50\% of the DR1 data release, and demonstrates the incredible fidelity and spatial dynamic range of the survey. The reduced data are available to the public at https://purcell.ssl.berkeley.edu/, and the raw data are also available upon request.

This paper begins by describing the overall signal chain (\S \ref{thesys}) and the observing techniques we employed in gathering the data (\S \ref{obstech}). We also describe the way the data were processed (\S \ref{redproc}), focusing on some of the new techniques we developed to process this unique data set, along with an example of the effect of the methods we employ in a sample region (\S \ref{exr}). We discuss the data presented in DR1 (\S \ref{dr1}), and outline our plans for the future of the ongoing GALFA-\hi survey (\S \ref{futdir}).

\begin{table*}[htdp]
\begin{center}
\begin{tabular}{c | c c c c c c }
survey  & telescope & era & area & resolution & velocity resolution & sensitivity \\
\hline
LAB & Dwingeloo and IAR & 1989 -- 2005 & 100\%&  $36^\prime$  &1.3  \kms &$\sim$100 mK  \\
IGPS & various & 1995 -- 2010 & 5\% & $1^\prime$--$2^\prime$ & 1.3 \kms & 2700 mK -- 1500 mK \\
GALFA-\hi & Arecibo & 2005 -- & 32\% & $4.0^\prime$ & 0.2 \kms & 60 mK -- 140 mK \\
GASS & Parkes & 2005 -- 2009 & 50\% & 15$^\prime$ & 1.0 \kms &60 mK  \\
EBHIS & Effelsberg & 2009 -- & $\sim$ 50\% & $9^\prime$ & 1.4 \kms &90 mK  \\
GASKAP & ASKAP & 2012 -- & 25\% &  $20^{\prime\prime}$ -- $3^\prime$ & 0.2 \kms & 3000 mK -- 40 mK  \\
\end{tabular}
\label{survcomp}
\caption{A comparison of the modern Galactic \hi surveys: LAB \citep{Kalberla05}, IGPS, (\citei{Taylor03}, \citei{Stil06}, and \citei{McC-G05}), GALFA-\hia, GASS \citep{McC-G09}, EBHIS \citep{Winkel10}, and GASKAP. Era describes the time period over which the data were taken and reduced. Area is shown in percent of entire sky covered. The IGPS data were taken at different resolutions and sensitivities by different sub-surveys. The GASKAP data can be reduced in different ways, leading to different resolutions and sensitivities. The sensitivity listed is for a 1 \kms channel.}
\end{center}
\end{table*}

%

\section{The System}\label{thesys}

\subsection{The AO 305 m}
The GALFA-\hi survey is conducted on the Arecibo Observatory 305-meter telescope, south of Arecibo, Puerto Rico, the central telescope of the National Astronomy and Ionosphere Center (NAIC). The primary reflector can be described as the 305 m diameter cap of a 265 m radius sphere. Suspended above the primary reflector is a triangular platform that carries a circular track upon which the azimuth arm rotates. The azimuth arm carries a geodetic dome (also called the Gregorian dome), which houses secondary and tertiary optics to correct for spherical aberration. The entire system can track objects from Declination -1$^\circ$ 20$^\prime$ to 38$^\circ$ 02$^\prime$ with $\sim 5^{\prime\prime}$ accuracy. ALFA is located on a rotating turret inside the dome at the focal plane of the telescope. The details of the 305 m are discussed in the Arecibo Technical and Operations Memos Series (ATOMS) and on the NAIC website, \texttt{http://www.naic.edu}.

\begin{figure*}
\begin{center}
\includegraphics[scale=.90, angle=0]{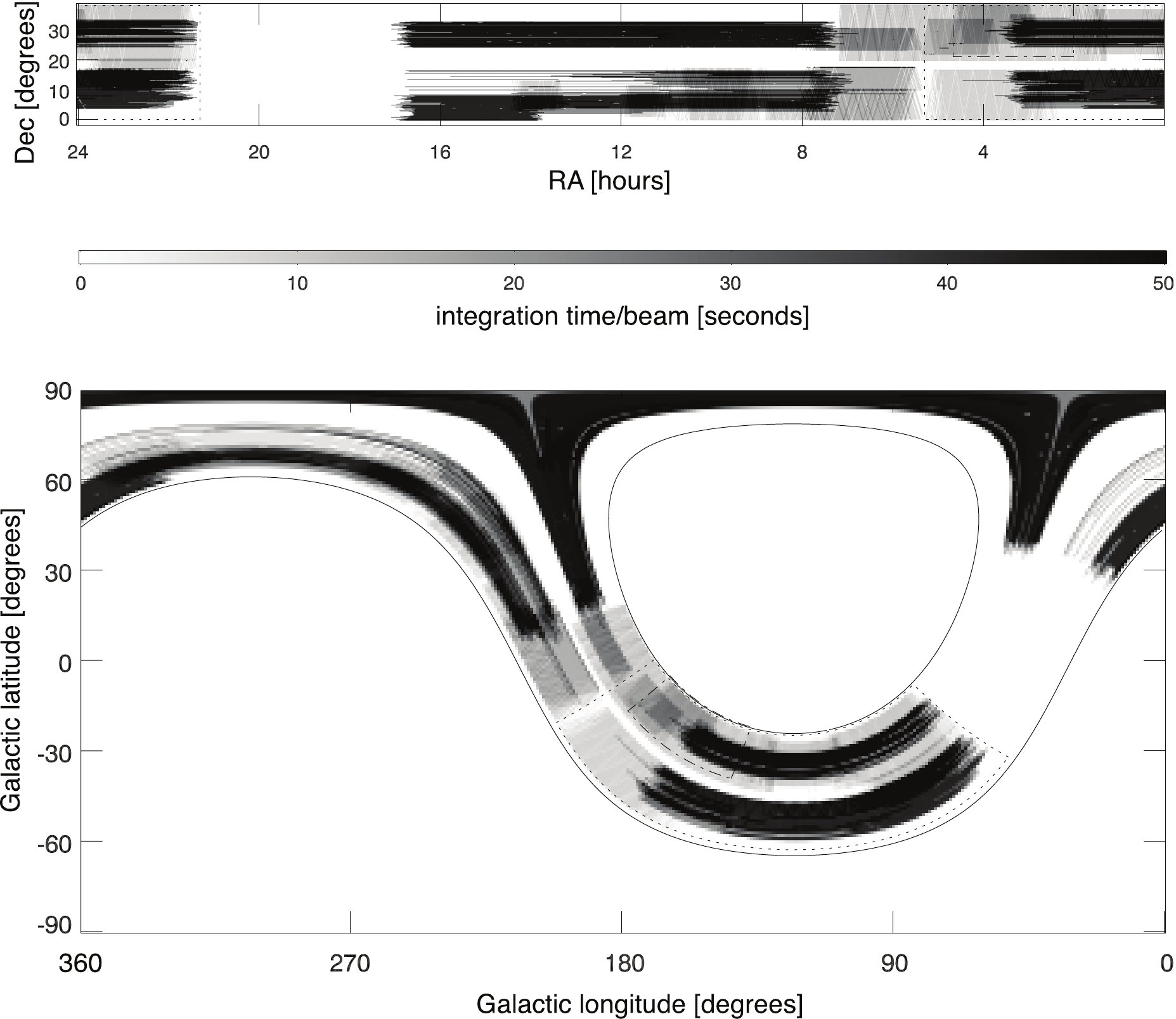}
\caption{The integration time of data release 1. The top panel is in equatorial coordinates, the bottom panel is in Galactic coordinates, with integration time per beam in grayscale. The rectangle in the top image indicates the total sky available to the Arecibo observatory, as does the black line in the bottom image. The dashed line represents the area shown in Figure \ref{pretty}; the dash-dotted line represents the sub-area shown in Figure \ref{slices}. \label{coverage}}
\end{center}
\end{figure*}

\begin{figure}
\begin{center}
\includegraphics[scale=.55, angle=0]{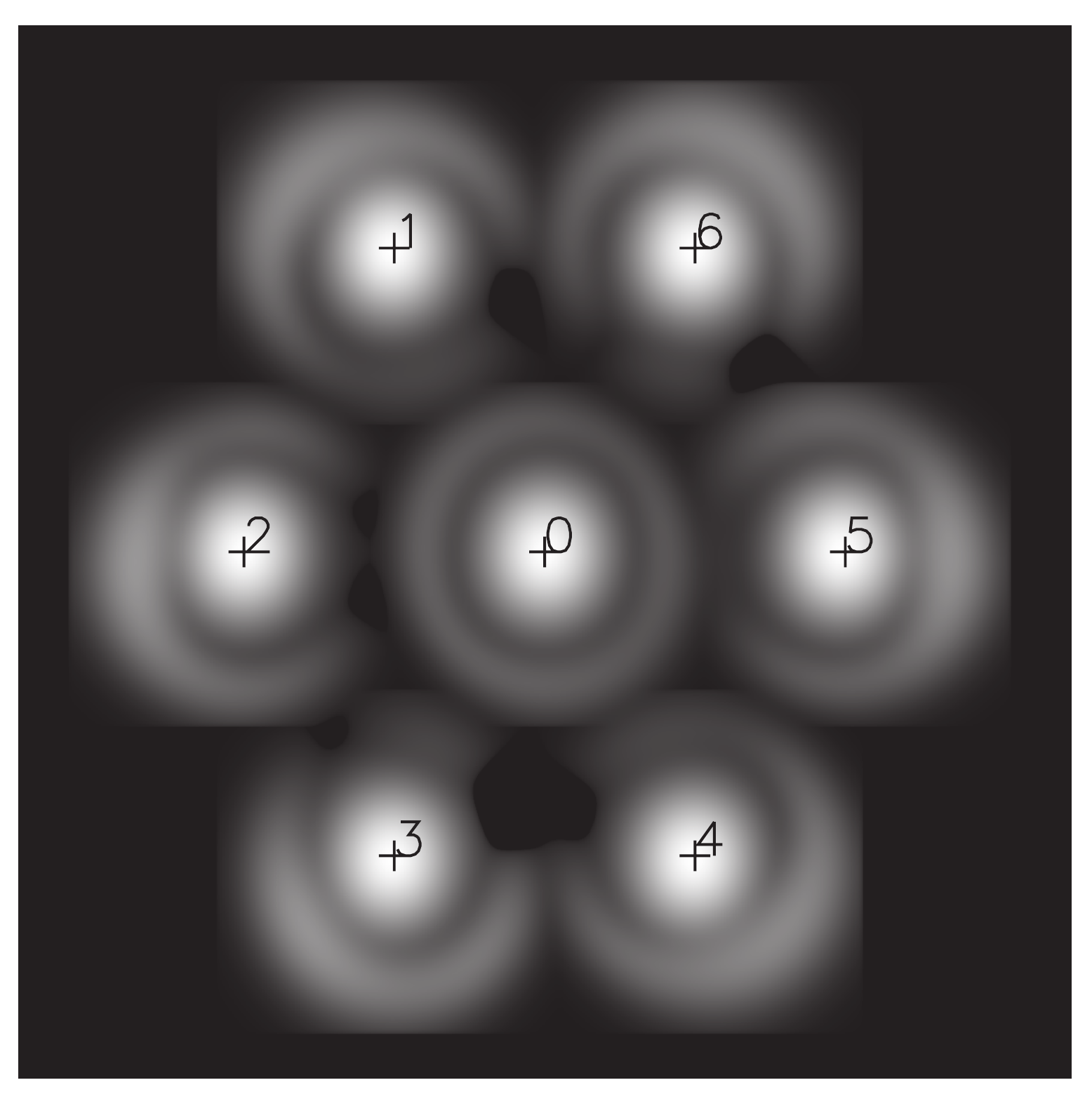}
\caption{A model of the response function of each of the 7 ALFA beams. Note the asymmetric sidelobe response in beams 1 -- 6. The beams are separated from each other in this diagram more than they are on the sky to show the sidelobes without collision.\label{slplot}}
\end{center}
\end{figure}

\begin{figure}
\begin{center}
\includegraphics[scale=.40, angle=0]{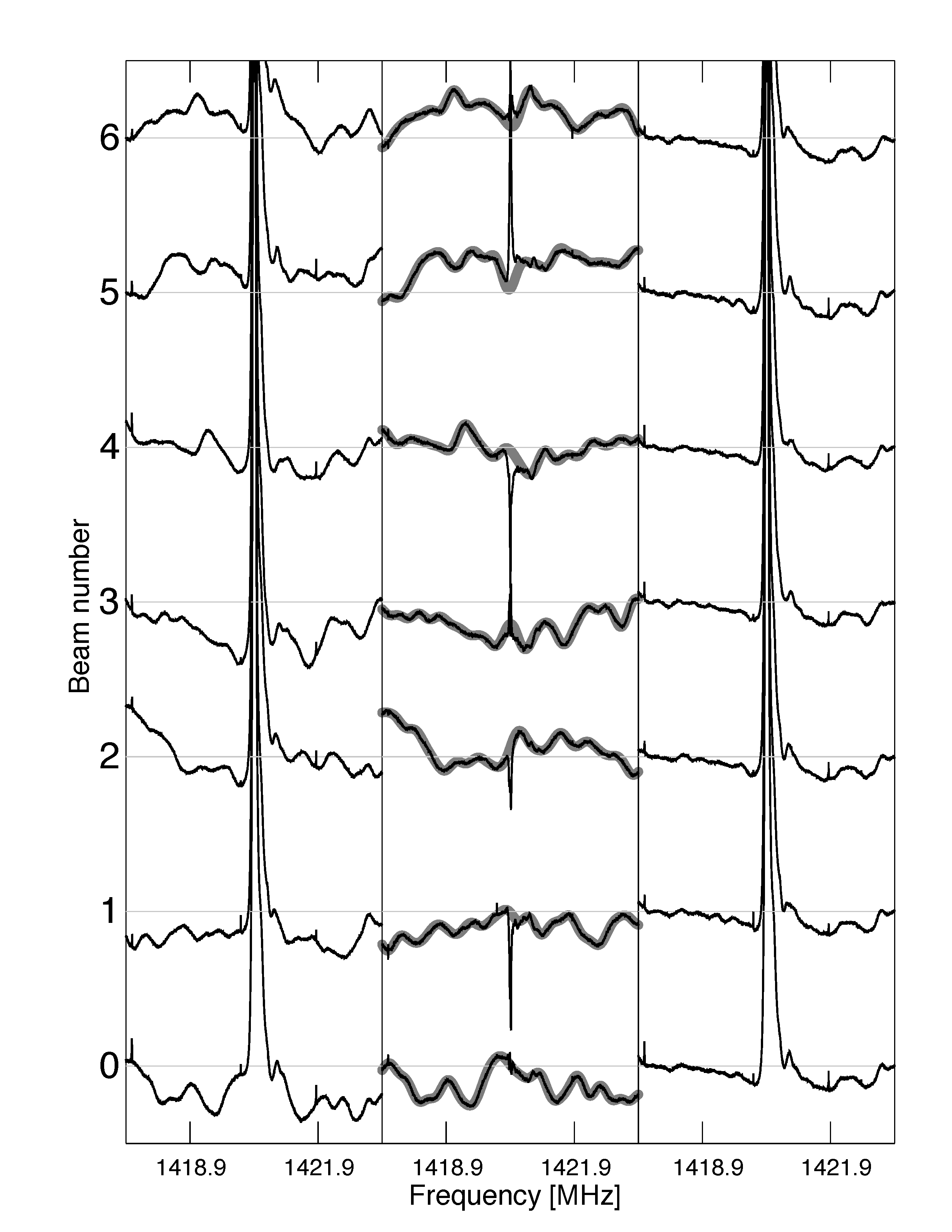}
\caption{An example of the FPN. The left panel shows the uncorrected FPN in each of the seven beams, averaged over polarization and over a few hours of observation. A gray line at zero is shown for comparison. The middle panel shows these same spectra less their average. Plotted behind these average spectra is the fit to these spectra in gray, as in the ripple component of Equation \ref{ripple}. The right panel shows the original spectra less the fit in the center panel. The ripple has been significantly reduced.\label{fpnex}}
\end{center}
\end{figure}

\subsection{ALFA}
ALFA is a 7-beam feed array designed to allow Arecibo to perform surveys at a 7-fold increased rate. The receiver frontend was built under contract to the Australia Telescope National Facility of the Australian Commonwealth Scientific and Industrial Research Organization, based upon the successes of the Parkes multi-beam receiver \citep{Staveley-Smith96}. The beams are arranged on the sky in a hexagon with a beam 0 in the middle (see Figure \ref{slplot}). Each beam is slightly elliptical along the zenith-angle direction, approximately $3.3^\prime \times 3.8^\prime$ full-width half-maximum (FWHM) on the sky, with a gain of $\sim$11 Jy/K for beam 0 and $\sim$8.5 Jy/K for beams 1-6. Each beam is divided into independent dual-linear polarizations, for a total of 14 independent signals. ALFA can be tuned from 1225 to 1525 MHz, which includes the hyperfine transition of \hi at 1420.406 MHz. For more design and implementation details consult the ALFA technical memo series, \texttt{http://www.naic.edu/alfa/memos/}.

\subsection{GALSPECT}
The GALFA-\hi spectrometer, GALSPECT, consists of 14 intermediate frequency (IF) to quadrature baseband downconverters, 7 dual polarization spectrometer boards, a CompactPCI computer Board, an Agilent 8648A synthesizer, all powered by an uninterruptible power supply. The system is designed to take 14 IF signals split and amplified from the ALFA signal chain and generate 14 32-bit spectra each second over 7679 channels covering $\sim$ 7 MHz as well as 14 32-bit spectra each second over 512 channels covering $\sim$ 100 MHz (see {\texttt http://seti.berkeley.edu/galfa} for details). The wide-band spectra are used to calibrate the narrow-band `science' spectra (see \S \ref{bfsfs}). The spectrometer has superb channel-to-channel isolation due to the use of a Fourier-transform polyphase filterbank and loses less than 1\% of data from `dead time' during data readout. The spectra are tagged with system parameters (telescope pointing, IF parameters, etc.) through the AO Shared Common Random Access Memory Network (SCRAMNet) and are written each second over the AO network to a large server for data storage and reduction.

\section{Observing Techniques}\label{obstech}
A suite of observing techniques is used in the GALFA-\hi survey, which we describe here. Many of the single-beam techniques these modes were based upon are described in \citet{ONeil02} and \citet{Stanimirovic02b}.

\subsection{GALFA-\hi Survey Philosophy}
When the GALFA-\hi project was started, the importance of both the legacy data product and target-oriented observing approach were recognized. These two goals could easily come into conflict if not treated carefully: too much freedom for observers to pursue individual projects could yield data that would be difficult to combine into a final survey product, too many constraints on the observations would make scientific goals hard to pursue. 
To facilitate the creation of a final survey, all of the GALFA-\hi data are taken in a single data-taking mode, with a fixed bandwidth, resolution and frequency setting. This is made possible by the GALSPECT spectrometer, which simultaneously has narrow enough channels (0.18 \kms) for the most stringent Galactic \hi science criteria (\hi narrow-line self-absorption, e.g., \citealt{Krco07}) as well as enough bandwidth (+700 to -700 \kms) to easily encompass the fastest known moving Galactic \hi gas (very high velocity clouds, e.g., \citealt{Peek07}). In this way all projects can be served by the same spectrometer setting. It is also important to allow individual observers to image higher-priority projects first. To observe the totality of the Arecibo sky, even at the shallowest possible level, would require more than 1700 hours of observing, thus delaying the publication of particular science of interest. The decision to allow individual PIs to choose science targets has indeed facilitated the timely release of data and the inclusion of students in the survey process \citep{Goldsmith03}. The DR1 coverage maps (Figure \ref{coverage}) show the effect of this philosophy, with each area having been observed for varying lengths of time and in varying observing modes (see Table \ref{dr1_proj}). As the survey continues, the entire Arecibo sky will be mapped, but at varying levels of sensitivity.

\begin{table*}[htdp]
\begin{center}
\begin{tabular}{c c | c c c c c c}
data set & principal investigator(s) & time & area & start date & end date \\
 & &hours & sq. deg. & MM/DD/YY & MM/DD/YY \\
\hline
\hline
{\bf DR1} &  &{\bf 3046} & {\bf 7520} & {\bf 05/16/05} & {\bf 02/26/09}\\
\hline
\hline
{\bf DR1-F} &  & {\bf 1429} & {\bf 4043} & {\bf 05/16/05} & {\bf 02/26/09}\\
\hline
a2032 & Stanimirovi\'c       &   41 &          203 & 05/16/05 & 09/03/05\\
a2011 & Peek \& Heiles &          13 &          111 & 05/21/05 & 06/12/05\\
a2050 & Peek \& Heiles &          41 &          414 & 06/25/05 & 01/21/06\\
a2004 & Paul Goldsmith &         38 &          337 & 07/09/05 & 12/31/07\\
TOGS & Putman \& Stanimirovi\'c  &         930 &         1766 & 08/17/05 & 01/14/09\\
a2172 & Heiles \& Peek &         106 &          974 & 06/07/06 & 02/18/07\\
a2222 & Peek &          63 &          615 & 07/29/06 & 01/17/07\\
a2174 &  Lewis Knee   &      33 &          316 & 09/07/06 & 11/05/06\\
TOGS2 &  Stanimirovi\'c      &  164 &         1526 & 11/14/08 & 02/26/09\\
\hline
{\bf DR1-S} & & {\bf1616} & {\bf3753} & {\bf06/17/05} & {\bf07/04/08}\\
\hline
a2011 &  Peek &         28 &          285 & 06/17/05 & 10/22/05\\
TOGS &  Putman \& Stanimirovi\'c &       1367 &         2672 & 12/10/05 & 05/31/08\\
a2221 & Leonidas Dedes & 10 &          104 & 08/19/06 & 09/17/06\\
a2220 & Heiles \& Peek &          40 &          429 & 10/28/06 & 12/04/06\\
a2060 &  Putman &        171 &          864 & 11/03/06 & 07/04/08\\
\hline

\end{tabular}
\caption{A table of parameters for DR1 and its subsets. DR1 is composed of two independently reduced subregions, DR1-S and DR1-F, all shown in bold. Each of these subregions are themselves composed of subregions, representing individual PI projects that have been combined together. They are indicated here by their Arecibo project number. The ``Turn on GALFA Survey'', a fully commensal set of observations that make up the lion's share of DR1, is indicated by TOGS (commensal with ALFALFA) and TOGS2 (commensal with GALFACTS).}\label{dr1_proj}
\end{center}
\end{table*}

\subsection{Least-Squares Frequency Switching}\label{lsfs}

In radio heterodyne spectroscopy, the measured spectrum is the product of the radio-frequency (RF) power and the intermediate-frequency (IF) gain spectra; to obtain the RF power spectrum, one must divide the on-source measured spectrum (the `on spectrum') by the IF gain spectrum. To accurately determine IF gain, we employ a new method called least-squares frequency switching (LSFS; \citealt{Heiles05b}). LSFS is fundamentally different from frequency switching. In frequency switching (the methodology used by GASS, for example; \citealt{McC-G09}), one takes two spectra---one on-line and one off-line
spectrum---and (1) finds the difference, to remove the continuum
(frequency-independent) system temperature, which is large; and (2)
divides the difference by the off-line spectrum, which corrects for the
frequency-dependent system gain. This technique introduces two
artifacts. First, the system temperature is not independent of
frequency, and in particular at Arecibo it has variations 
on the $\sim 1$ MHz scale, known as fixed-pattern noise (FPN; see \S \ref{ripsub} and Figure \ref{fpnex}). Thus, when taking the difference in (1), the FPN in both the on-line and the off-line spectra appears in the
difference spectrum; these are uncorrelated, so the FPN amplitude
increases by $\sim \sqrt{2}$. Second, the system gain depends mainly on
IF frequency, and the division in (2) reliably corrects for this
IF gain component. However, the system gain also varies with RF	
frequency to some extent, and (as with the FPN) RF gain
variations in the off-line spectrum appear in the final result. Our LSFS technique
avoids some of these problems because it avoids the need for
subtracting an off-line spectrum, and it derives the IF gain
alone, so that the final derived spectral gain errors come only from
those in the RF spectrum.

In LSFS, one sets the local oscillator (LO) frequency
to a number of different values so that the RF power spectrum and the IF
gain spectrum can be evaluated as distinct entities using a
least-squares technique \citep{Heiles07}. In practice, at the beginning
of each observation we spend 10 minutes recording data at a single sky
position at a number of different frequencies. We used the
minimum-redundancy schema with 7 LO settings. In this ``MR7'' schema,
the LO frequency separations range from 1 to 39 times the minimum
separation $\Delta f$. One needs the maximum separation to be a
reasonable fraction of the total spectrometer bandwidth so that the
full-bandwidth gain spectrum is accurately defined. With our large
number of channels, this means that $\Delta f$ must be much larger than
the channel spacing $\delta f$. We chose ${\Delta f \over \delta f} =
225$, which gave a maximum separation of 90\% of the full spectrometer
bandwidth.

When ${\Delta f \over \delta f} > 1$, one needs a reliable way to
degrade the resolution of the original 7679-channel spectra to reduce
the number of channels in the LSFS technique to a manageable
number. One applies the LSFS technique to these degraded-resolution
spectra to derive a degraded-resolution IF gain spectrum. Finally, one
interpolates the degraded-resolution gain spectrum to regain the
original number of spectral channels. We used standard Fourier series to
both degrade the original resolution and to then recover it, reducing the number of channels in the degraded-resolution
spectra to 512.

\subsection{Scanning Methods}

	\subsubsection{Drift Scanning}
\begin{figure}
\begin{center}
\includegraphics[scale=.44, angle=0]{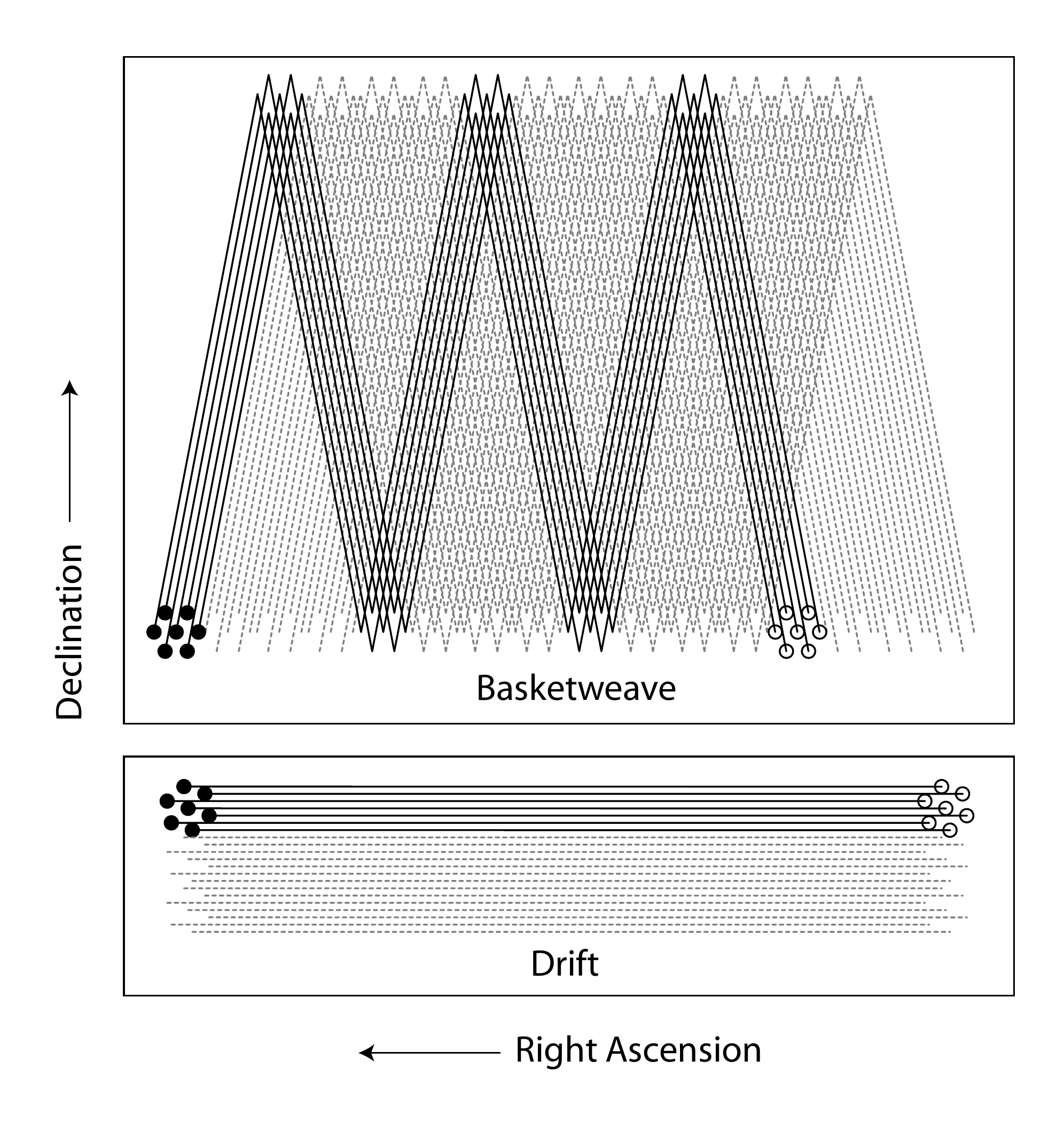}
\caption{The two modes of scanning used by GALFA-\hi Data Release 1. In the examples, the empty circles represent the initial position of ALFA for a single day's observation and the filled circles represent the final position of ALFA for the day's observation. The dashed lines represent complementary scans taken on other days of observations.
\label{scanpattern}}
\end{center}
\end{figure}

The simplest method of observing is the drift-scanning method. Indeed, this method traces back to the dawn of radio astronomy as the telescope is simply left at a fixed position at transit (in our case at azimuth = 180 or azimuth = 0) and the sky is allowed to drift by \citep{Reber40}. To space the beams equally ($\sim 2^\prime$) the ALFA beams are rotated $\sim 19^\circ$ from the orientation shown in Figure \ref{slplot}, as shown in Figure \ref{scanpattern}. This scanning pattern holds the maximum number of telescope parameters fixed, which minimizes variability in baseline ripple. This pattern also has a number of disadvantages. Firstly, it is unwieldy for regions that are relatively large in Declination (or small in Right Ascension). Secondly, since the scans are all parallel on the sky, there is no easy way to cross-calibrate the scans, as they do not intersect (see \S \ref{bw}). Thirdly, the telescope has limitations on the zenith angle at which it can point, so regions near zenith are inaccessible in the standard mode. It is possible to drift scan in positions off transit to get Declination range near zenith, but one sacrifices constancy of FPN. This method is used solely by the Arecibo Legacy Fast ALFA Survey (ALFALFA; \citealt{Giovanelli05}), which is a wide-area survey with which GALFA-\hi commensally observes. Data taken commensally with ALFALFA (the Turn On GALFA Survey, or TOGS, project) represent 75\% of the raw data that make up DR1, and cover 60\% of the DR1 sky. This means that the TOGS areas are some of the highest sensitivity areas of the DR1 data set (see the dark bands at fixed declination in Figure \ref{coverage}).

	\subsubsection{Basketweave Scanning}\label{bw}
The preferred mode for GALFA-\hi observations is basketweave or `meridian-nodding' scanning. In this mode the telescope is kept at the meridian and is driven up and down in zenith angle over the chosen Declination range. Each day the starting position of these nods is displaced in Right Ascension by $\sim 12^\prime$ such that the entire region is covered, as in Figure \ref{scanpattern}. Alternatively, the scan can be started at a fixed RA, and the Declination can be varied, which is the method used by our commensal partner GALFACTS (\citei{GT09}). There are six distinct drive speeds (or `gears') that generate equally spaced tracks with each of the seven beams: three with an ALFA rotation angle of $0^\circ$ and three with an ALFA rotation angle of $30^\circ$. Practically, only the two fastest gears, one with rotation angle $0^\circ$ and one rotation angle of $30^\circ$, are used. This observing method has the advantage that it generates a large number of intersection points among beam tracks at which the relative gain of the beams can be measured (as in \citealt{Haslam81}). This allows us to greatly reduce the striation in the map caused by gain variation. It does this while maintaining a fixed azimuth, which significantly limits the fluctuations in FPN. 

A somewhat modified version of the basketweave can be employed at Declinations that pass near zenith at Arecibo where the traditional basketweave scan cannot be used.  In this method, called `azimuth wagging', the same scan pattern on the sky is achieved, but by positioning the telescope at near 90 or 270 degrees azimuth. This method is inferior to the standard basketweave in that the azimuth, zenith angle and ALFA rotation angle must all change continuously during the observation.

\section{The Reduction Process}\label{redproc}

\subsection{Bandpass Fitting with LSFS}\label{bfsfs}
As discussed in \S \ref{lsfs}, we need to correct each spectrum for the IF gain.  At the beginning of each observing period, we run an LSFS calibration to obtain this IF gain spectrum and we use this spectrum for the whole period's data. This works because the IF gain spectrum is largely stable. We find that from day to day the IF solutions typically vary with amplitudes of only dozens of mK, but larger jumps of up a few K are observed occasionally. We attempt to maximize this stability by setting all aspects of the IF electronics, including all attenuator and amplifier gain values, to standard values. Occasionally LSFS data are contaminated, either by RFI or by improper implementation during the observation. In this case we use the LSFS data product from the previous or following day of observation.

	\subsection{Initial Calibrations}\label{initcal}
\begin{figure*}
\begin{center}
\includegraphics[scale=.50, angle=0]{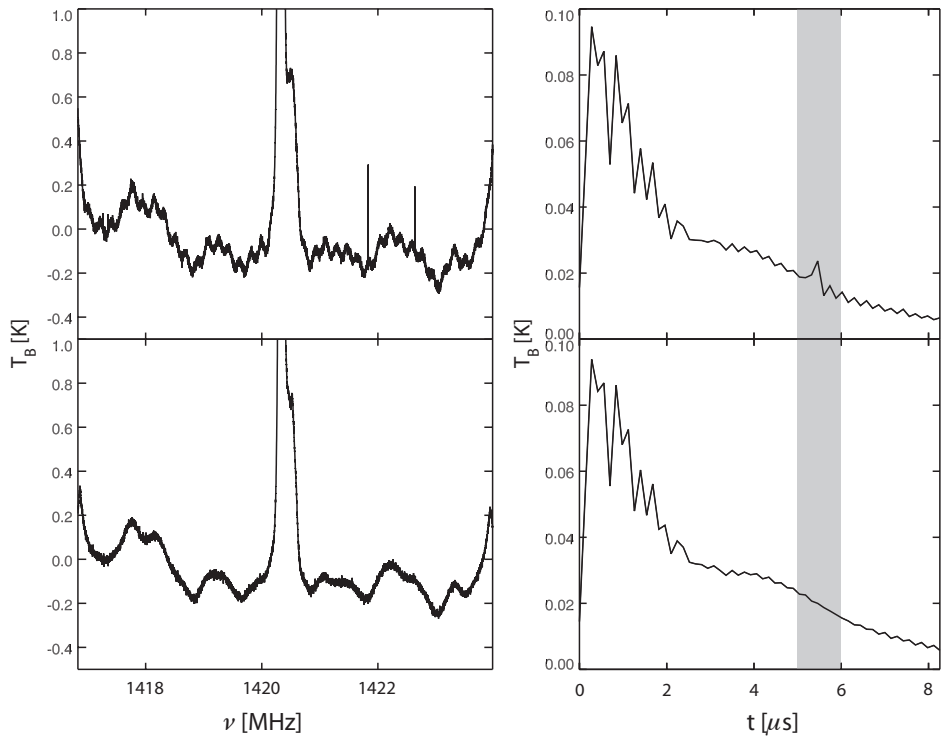}
\caption{The plots on the left are time averaged spectra from a drift scan observation in the original state (top) and with the cable reflection removed (bottom). The plots to the right are the Fourier transform of the spectra, with the region where the reflection was removed highlighted. Note that the digital Fourier transform of a monochrome signal in general has both a central spike and ``ringing'' component in Fourier space; our fitting also largely reduces the ringing component. The longer period and more chaotic FPN has not been reduced in this example. \label{refl_ex}}
\end{center}
\end{figure*}

After the effects of the bandpass are removed, the data are searched for possible contamination by reflections in the signal chain (see Figure \ref{refl_ex}). Any impedance mismatch between cables or signal transmitting hardware can cause echoes within the system. Similar to the reflections in the superstructure and geodetic dome of the telescope, these reflections in the cables produce excess power at lags equivalent to the reflection's differential travel time, and therefore cause ripples in the baseline of the spectrum. These reflections have the very nice quality that they are each only a single Fourier mode: as the reflection geometry is only one dimensional, there is no room for complex reflection patterns as there is in the telescope superstructure. This ripple is therefore very easily removed by examining the data in Fourier space and simply removing the clear reflection tone. We also remove any single-channel RFI that is detectable above the noise in the data smoothed over a ten-second interval by setting it to the median of the adjacent channels. A secondary RFI effect is also searched for: radar intermodulation products. In certain LO configurations, a periodic radar signal emanating from the San Juan airport can generate periodic pulses in our data that can be hundreds of channels wide and as much as 10 K in amplitude. These pulses have a very well known period and spectral shape, so we can search for them and eliminate them from the data relatively easily. After these spectral cleaning procedures are done the data are shifted into the local standard of rest (LSR) frame.

	\subsection{Fixed-Pattern Noise Subtraction}\label{ripsub}
	
It was known since long before the GALFA-\hi project was undertaken that spectra taken at the Arecibo 305 m (and all radio telescopes) are typically contaminated by fixed pattern noise (FPN; see, for example, \citealt{Padman74} and references therein). FPN (also called  `baseline ripple') at Arecibo manifests as a low-level wave that permeates the entire spectrum, with peak-to-trough variations of $\sim 400$ mK across a single, polarization-averaged band (see the first panel in Figure \ref{fpnex}; for a detailed description of ALFA's FPN see \cite{Heiles05}). Note that while this is certainly strong FPN, it does not dramatically exceed values found for telescopes with completely unblocked apertures; the Green Bank Telescope has FPN at about 30\% this level in L-band \citep{FNB03}. At Arecibo, FPN contamination to the spectra is caused by reflections inside the Arecibo 305 m superstructure and geodetic dome. Any signal that takes more than a single path to the receivers will cause a spurious increase in the amplitude of the auto-correlation function (ACF) in the channel corresponding to the delay between the paths. In the case of a signal that takes both a standard path to the receivers and one that has an extra reflection between the geodetic dome and the main reflector, for instance, the path length difference is on the order of 300 meters, which corresponds to a delay of 1 $\mu$s. This 1 $\mu$s delay causes an increased ACF at the 1 $\mu$s lag which, in turn, generates a ripple with a period across the spectrum of 1 MHz, or 200 \kmsa. Since Galactic \hi typically covers a comparable velocity range, this FPN is capable of corrupting the final data product if not addressed adequately. 


Removing the FPN caused by the reflections in the telescope superstructure and geodetic dome is much more difficult than removing the effects of reflections within the signal chain. This is because the FPN is not dominated by a single Fourier component, but rather by a large range of Fourier modes, with associated delay ranging from 0.5 to 2 $\mu$s. The FPN was examined in some detail in \citet{Heiles05}. This work showed that the ripples are 100\% polarized, uncorrelated from receiver to receiver, relatively static over time and dependent upon ALFA rotation angle, telescope azimuth and zenith angle. 

To reduce the effects of the FPN we use the fact that each beam has an uncorrelated and relatively fixed pattern over the course of an observation in our two main observing modes, drift scanning and basketweave scanning. In the case of drift scanning all telescope position parameters are held fixed, so we expect no variation in the FPN. In the case of basketweave scanning only the zenith angle is changed, which causes an average change of $\sim 25 \%$ in the FPN over the full range of zenith angle, which we ignore. After averaging the two polarizations within a beam and applying a simple overall gain calibration, we reduce the effect of the FPN by first generating the average spectrum, $T_{avg}\left(\nu \right)$ for all 7 beams over the course of a single day's observation. We then fit the difference between $T_{avg}\left(\nu \right)$, and the average over beam $n$, $T_{n}\left(\nu \right)$, with 
\begin{eqnarray}\label{ripple}
&& T_{n}\left(\nu \right) - T_{avg}\left(\nu \right) = \sum_{i=1}^m \left( a_{i}sin\left(t_i\nu\right) + b_icos\left(t_i\nu\right)\right)  \\
&&+ \sum_{j=1}^l \left(\left(\frac{\partial T}{\partial\alpha}\right)_{\nu_j}\Delta\alpha_nd\left(\nu - \nu_j\right) + \left(\frac{\partial T}{\partial\delta}\right)_{\nu_j}\Delta\delta_nd\left(\nu - \nu_j\right)\right) \nonumber
\end{eqnarray}
where, $t_i$ ranges over the observed FPN frequencies (0.5-2 $\mu$s) and $\nu_j$ ranges over the frequencies where Galactic disk \hi is typically observed. $d$ is equivalent to the Dirac delta function, to avoid confusion with Declination, $\delta$. The first component of the fit is the baseline ripple, the second component is the residual \hi contribution. By fitting the residual \hi contribution with this Taylor expansion in $\alpha$ and $\delta$ we are able to accurately capture the \hi residuals while only fitting 2 parameters per channel, rather than the 6 required if we were to fit each beam individually. By subtracting only the baseline ripple component of the fit for each beam, we are able to reduce the amplitude of the baseline ripple in each beam by $\sim \sqrt{7}$ without disturbing the \hi data. An example of this method is shown in Figure \ref{fpnex}.

	\subsection{Intersection-Point Calibration}\label{intpnt}
	
\begin{figure}
\begin{center}
\includegraphics[scale=.30, angle=0]{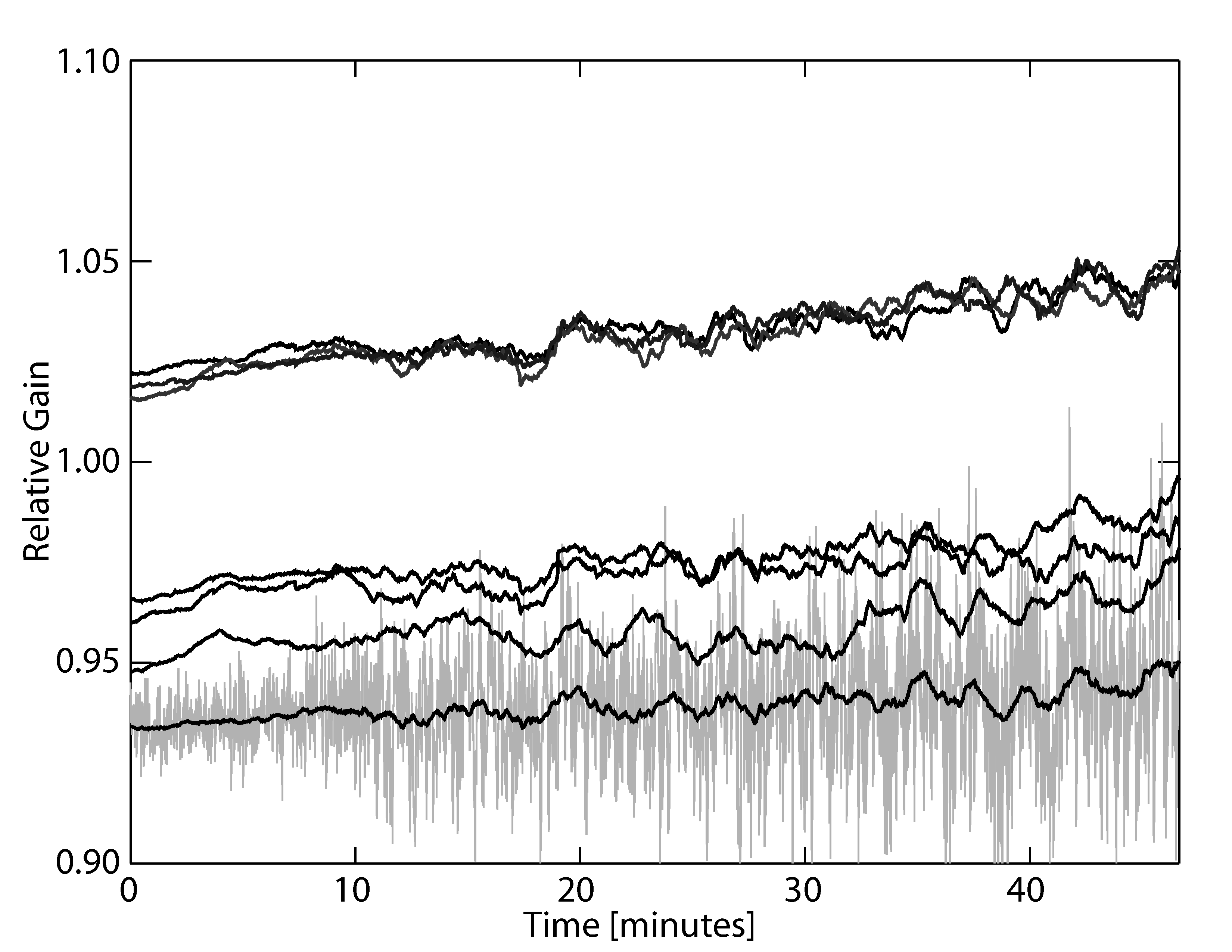}
\caption{The relative gains between two consecutive days of observation, taken at identical positions. Each coincident integration is compared for each polarization-averaged beam, plotted in black (one line for each beam). We find the relative beam gain between the two days at each second by fitting to the \hi spectrum taken on the first day plotted against the \hi spectrum taken on the second day, where the relative gain is the slope of the line. Each relative beam gain is averaged over 100 seconds to smooth the gain trends. The smoothed beam 0 gain data are plotted in black, with the unsmoothed beam 0 relative gains in light gray for comparison. The noise in the unsmoothed gain fits changes over the course of the observation time as the observations leave the Galactic disk and the \hi amplitude decreases significantly.\label{gainex}}
\end{center}
\end{figure}

	To make a correct map from data taken with a single dish radio telescope, the overall gains need to be accurately calibrated. These gains can vary significantly from day to day, based on both the performance of the telescope and the receiver (see Figure \ref{gainex}, which shows the variation on day and minute timescales). These gains may even vary during long observations. It is possible to calibrate these variable gains with the calibration noise diodes in the ALFA system, but the amplitude at the receiver that these diodes generate can also vary significantly from day to day. Uncalibrated gains lead to striation in the maps and inaccurate \hi amplitudes. 

To determine the relative gains of each beam over each day of observation, we use the constraint that at places where beam tracks intersect they must observe the same region of sky, so any differences in their spectra must be dominated by variation in the gain parameters. This method is most effective in basketweave scans where there are many intersection points -- a region 2 hours across in Right Ascension and 17 degrees in Declination has almost $10^5$ intersection points. If we wish to determine the relative gains in a set of drift scans, we must conduct other observations that intersect the drift scans, as drift scans are parallel on the sky.

Once the intersection points are found, we determine the relative gains at these points by comparing the amplitudes of the observed \hi spectrum in each beam. Each beam is assumed to have an overall gain for each day's observation plus some variability with time during the observation that we parameterize as a sum of Fourier components. Since only relative gains are measured, the overall gain solution must average to unity. We construct a least-squares fit to the measured relative gains at intersection points for the various day and time dependent gain parameters and then apply these parameters to the entire data set to remove the relative gains.

	\subsection{Gridding}\label{grid}
Once the spectra, or time-ordered data (TOD), have been fully calibrated, we apply them to a grid in the image plane. This grid can be made in any projection in the World Coordinate System as described by \citet{CG2002}, but we project to Cartesian coordinates in Right Ascension and Declination with an orthogonal graticule. We use a Gaussian sampling kernel with FWHM = $3^\prime$ to place flux on the grid; this size sampling kernel smoothes out any pixels that may have been missed by the observing pattern without degrading the resolution too significantly. Our method is anchored by the IDL code SDGRID.PRO, a code based upon the IDL gridzilla implementation by Sne\v zana Stanimirovi\'c. SDGRID.PRO uses the core logic described in \citet{Dickey02} and many of the methods developed for Parkes gridzilla \citep{Barnes01} and the AIPS procedure SDGRID. At this point the first sidelobe correction is applied (see \S \ref{fsc}, below). Once sidelobe correction is complete, a final amplitude correction is applied to make the observations consistent with the lower resolution Leiden-Argentine-Bonn Survey \citep{Kalberla05}. We select a large region of sky towards the plane, where stray radiation will be relatively weak, and compare the GALFA-\hi spectrum to the LAB spectrum for the region. This comparison results in a 10\% to 15\% adjustment of the final GALFA-\hi data cubes.
	
\subsection{First Sidelobe Calibration}\label{fsc}
	
Each ALFA beam has a significantly different beam shape, because of the very fast optics of the telescope \citep{Heiles04}. The center beam (beam 0) has a well behaved beam shape, with sidelobe efficiency, $\eta_{SL}$, near 10\%.
Each other beam (beams 1 -- 6) has significant first and second sidelobes which depend upon the relative displacement of the beam to beam 0 (see Figure \ref{slplot}). The first sidelobes are $\sim 5^\prime$ offset from the main beams with $\eta_{SL}$ near 20\% for beams 1 to 6. Since we do not Nyquist sample the sky with each beam, but rather with a combination of all the beams, we end up with a map that has an ill-defined associated beam pattern. This means that many of the standard methods devised for removing the effect of beam shape from maps, such as the maximum entropy method (\citei{CB89} and references therein) and CLEAN \citep{Hogbom74}, cannot be used and we must devise our own method for removing the effects of these sidelobes. We note that the GALFACTS survey has developed a `multi-beam CLEAN' algorithm \citep{GT09}, though it is primarily designed for point-source reconstruction, rather than the diffuse Galactic \hi signal we detect.

To counteract the effect of these inconsistent, asymmetrical beams we use a strategy similar to that employed to correct the Leiden-Dwingaloo Survey, as outlined in \citet{HKBM96}. Because we have seven different beams, rather than a single convolving beam, we must modify their method somewhat. The method, in essence, is to determine a ``contamination spectrum'' for each of the TOD; these contamination spectra represent what we expect the sidelobes alone would detect. We construct these contamination spectra by `re-observing' the data cube at each of the locations of the TOD, assuming the sky is identical to our data cube from \S \ref{grid}. These contamination spectra are then subtracted from the original spectra,
\beq
T^\prime\left(\nu, t\right) = T\left(\nu, t\right)/\eta_{\rm MB} - \eta_{\rm SL}T_{C}\left(\nu,t\right)/\eta_{\rm MB},
\eeq
Where $T\left(\nu,t\right)$ are the original TOD, $T_{C}\left(\nu,t\right)$ are the contamination spectra, and $\eta_{\rm MB}$ and $\eta_{\rm SL}$ are the main beam and sidelobe efficiencies, respectively. These corrected spectra, $T^\prime\left(\nu, t\right)$ are then re-gridded to make a final map. While there are second-order effects that stem from using the contaminated map as the true sky image, we have found that these effects are small, of order $\eta_{\rm SL}^2$, and typically smaller than the effect from unmodeled higher order sidelobes.

We measure the effective beamsize of the final maps by measuring the size of compact radio sources in \hi absorption. We examine the 100 brightest NVSS sources that are both within DR1 and in regions of the sky where the Galactic \hi does not vary significantly over the region near the source. We find that an effective beamsize of 4.1$^\prime$ in Declination and 3.9$^\prime$ in Right Ascension. This ellipsoidal nature of the effective beam is due to the known elongation of the Arecibo beam along the zenith angle axis (e.\ g.\ \citealt{Heiles04}), which is kept oriented along Declination during our observations. 

\section{Data Processing: An Example Region}\label{exr}
As an example of the GALFA-\hi survey observations and data processing methods we have described in the previous sections, we show here a specific region observed over several months in 2006. The entire region is $20^\circ \times 17^\circ$, centered on $(\alpha,\delta)=(17^{\rm h},+8^\circ)$, and was chosen by the investigators (Putman and Stanimirovi\'c) because HVC complex C is visible in the region. It was observed with the basketweave method (see \S \ref{bw}) twice, to achieve a higher signal-to-noise. 

\begin{figure}
\begin{center}
\includegraphics[scale=.40, angle=0]{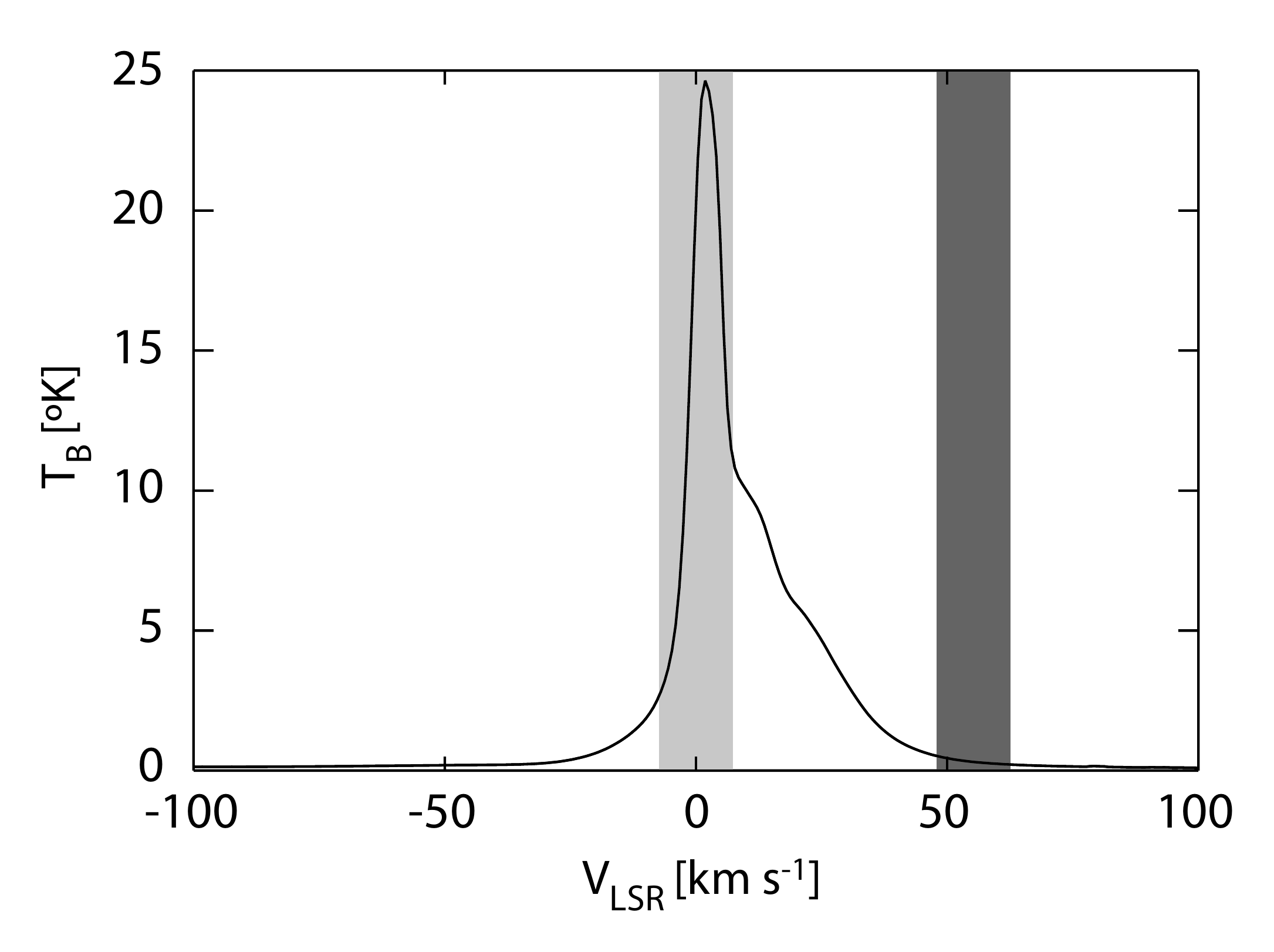}
\caption{The average spectrum over the region discussed in \S \ref{exr}. The on-line and off-line maps were integrated over the velocity ranges indicated in light gray and dark gray domains, respectively. \label{ex_spec}}
\end{center}
\end{figure}

\begin{figure*}
\begin{center}
\includegraphics[scale=.75, angle=0]{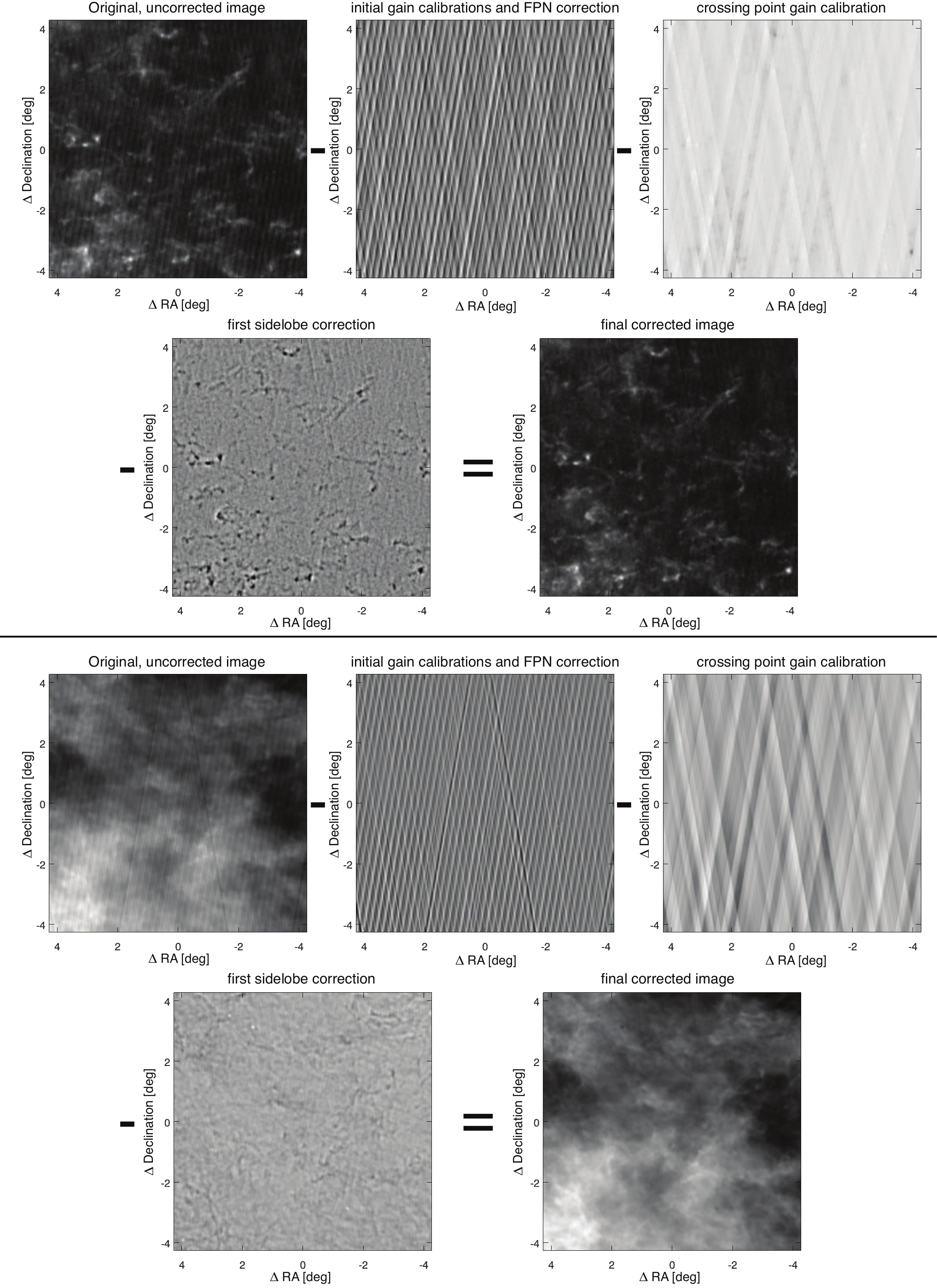}
\caption{An example region of the sky towards $(\alpha,\delta)=(17^{\rm h}\,52^{\rm m}, +10^\circ\, 21^\prime)$. The top and bottom areas represent the dark (off-line) and light (on-line) regions in Figure \ref{ex_spec}, respectively. The maps are scaled to fit the range of observed surface brightnesses, from 0 K to 2 K in the case of off-line map (top) and from 7 K to 23 K in the case of the on-line map (bottom). The calibration images are scaled from -0.1 K to 0.1 K in the top panels, and from -1 K to 1 K on the bottom panels. See \S \ref{exr} for details. \label{ex_reg}}
\end{center}
\end{figure*}

In this example we investigate a single survey cube, which is $512 \times 512~ 1^\prime$ square pixels and in the standard Cartesian projection and 2048 frequency channels. Figure \ref{ex_spec} shows the average spectrum over the cube and the two velocity ranges we consider. We isolate these velocity ranges to demonstrate the effect of each stage of the data reduction on both strong \hi (on-line) and weak \hi (off-line) maps. We demonstrate the effects of each stage of reduction in Figure \ref{ex_reg}, by showing a map of the uncalibrated data at the top left, fully-calibrated data at the bottom right, and the corrections applied at each step in between. We set these up like an equation, with the final data cube equalling the original data cube less each isolated contaminant (``initial calibrations and FPN'', ``crossing point gain calibrations'', and ``first sidelobe corrections''). The maps are scaled to fit the range of observed surface brightnesses, from 0 K to 2 K in the case of off-line map (top) and from 7 K to 23 K in the case of the on-line map (bottom). Correction maps are scaled from -0.1 to 0.1 K (top 3 correction maps) and -1 K to 1 K (bottom 3 correction maps). Note that the software does not typically generate these intermediate images, but we show them here for clarity. 

The first two panels (``original, uncorrected image'') in the top and bottom section of Figure \ref{ex_reg} show the integrated intensity map of the off-line and on-line regions without any calibration past the initial calibrations described in \S \ref{initcal}. The strong striation along the scan patterns from poorly calibrated gains and baseline ripple are quite evident. The following three pairs of plots show the initial gain calibrations and FPN correction (\S \ref{ripsub}), crossing-point gain calibration (\S \ref{intpnt}), and the first-sidelobe correction (\S \ref{fsc}), respectively. For the first (FPN) correction, there is a strong effect in both the on- and off-line data. In the case of the off-line plot (top), the cleaning is primarily due to the ripple reduction itself, as the amplitude of the features on the sky is similar to the amplitude of the ripple. In the on-line slice (bottom) the cleaning is not due to baseline ripple, but rather the fact that a rough gain calibration applied when the baseline ripple is corrected. In the second pair, ``crossing point gain calibration'', there is only a modest change due to gains, whereas the much brighter on-line data show significant changes in some regions. The final step of first-sidelobe correction has most effect at the bright peaks of the slices where data had been erroneously scattered, and has a noticeable effect on both off-line and on-line slices. The last pair of maps shows the same region and velocity range as the first set of maps, but with all three data reduction methods applied. The off-line data are significantly improved, but given the low dynamic range of the map ($\sim 1$ K), it is not surprising that residual ripple on the order of 0.1 K (see Figure \ref{fpnex}) is still visible. The on-line data are also drastically improved, and even with very close inspection it is very hard to determine the presence of any systematic artifacts. These maps demonstrate the very high quality of the GALFA-\hi data products.

\section{Data Release 1}\label{dr1}

\begin{figure*}
\begin{center}
\includegraphics[scale=.60, angle=0]{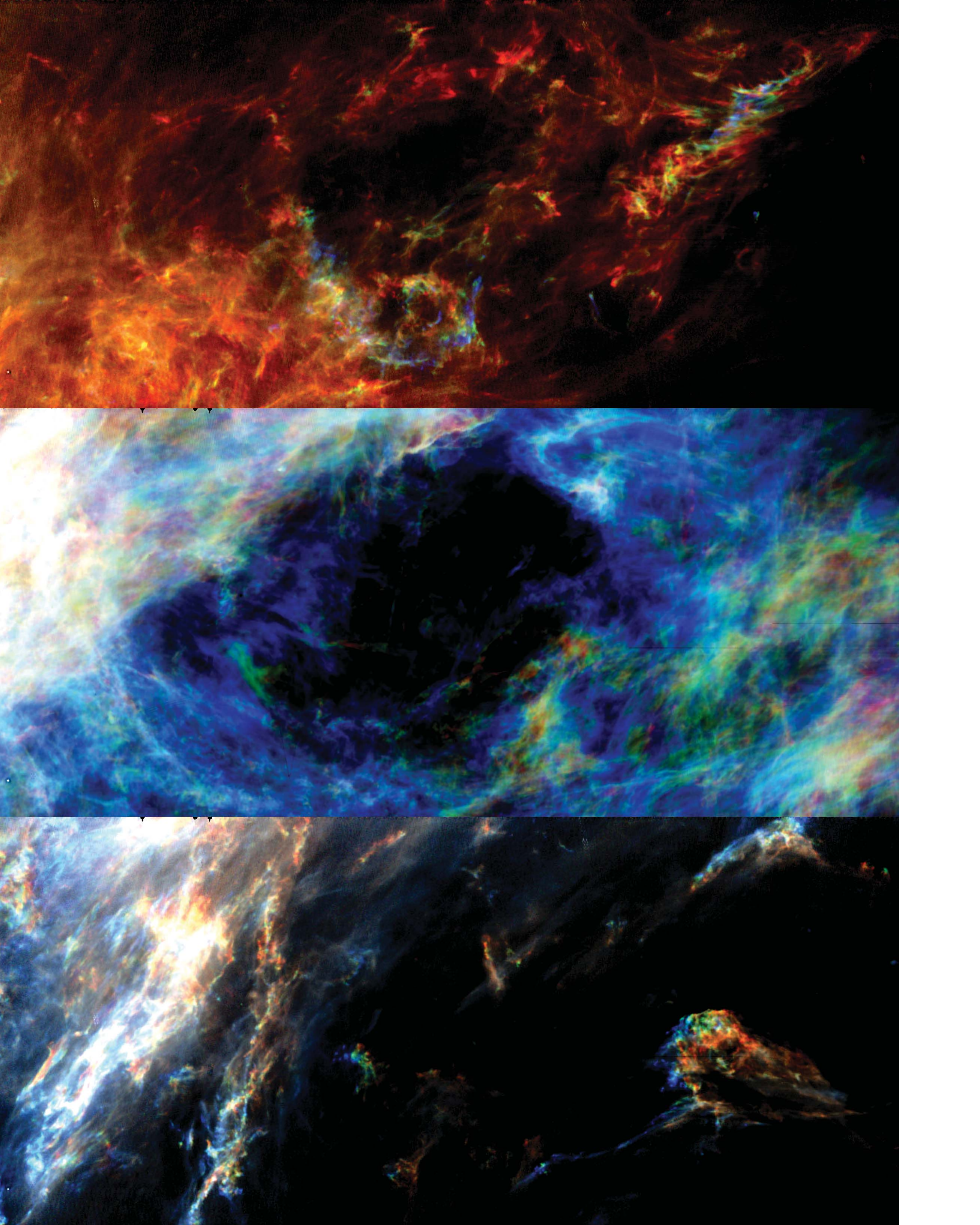}
\caption{An typical region of the sky towards $(\alpha,\delta)=(3^{\rm h}\,20^{\rm m}, +29^\circ\, 55^\prime)$, 40$^\circ \times18^\circ$ $10^\prime$ in size. The top panel represents, -41.6, -39.4, -37.2 \kms in red, green, and blue, respectively. The middle panel represents -4.0, -1.8, and 0.4 \kmsa, while the bottom panel represents 15.8, 18.7, 21.7 \kmsa. \label{slices}}
\end{center}
\end{figure*}

Data Release 1 (DR1) is the first full data release of the GALFA-\hi survey. The 3046 hours of data from 12 independent projects were taken from May 2005 to February 2009: details can be found in Table \ref{dr1_proj}. Integration times per beam area range from $\sim$10 seconds to $\sim$60 seconds (see Figure \ref{coverage}), as different regions were observed in different modes, and sometimes covered by more than one project. These integration times correspond to 140 mK and 60 mK RMS noise, respectively, when smoothed to a 1 \kms channel width. The median integration time per beam is 30 seconds (80 mK). DR1 was reduced in two separate stages; DR1-S (spring), covering the sky towards 12h RA, and DR1-F (fall), covering the sky towards 0h RA. DR1-S was made publicly available in January 2009, and the entirety of DR1 is released with this work. The two areas have an overlap of 180 square degrees toward $(\alpha,\delta)=(6^{\rm h}, 25^{\circ})$, and give consistent results. In Figure \ref{slices}, we demonstrate the quality of the DR1 data set, and the fantastic structure of the ISM at 4$^\prime$ resolution, by showing three sections of velocity space over the same arbitrary chosen 40$^\circ \times18^\circ$ $10^\prime$ region. The intense filamentary and diversity of cloud structures is immediately apparent. 

\subsection{DR1 data reduction and known systematics}

\begin{figure}
\begin{center}
\includegraphics[scale=0.55, angle=0]{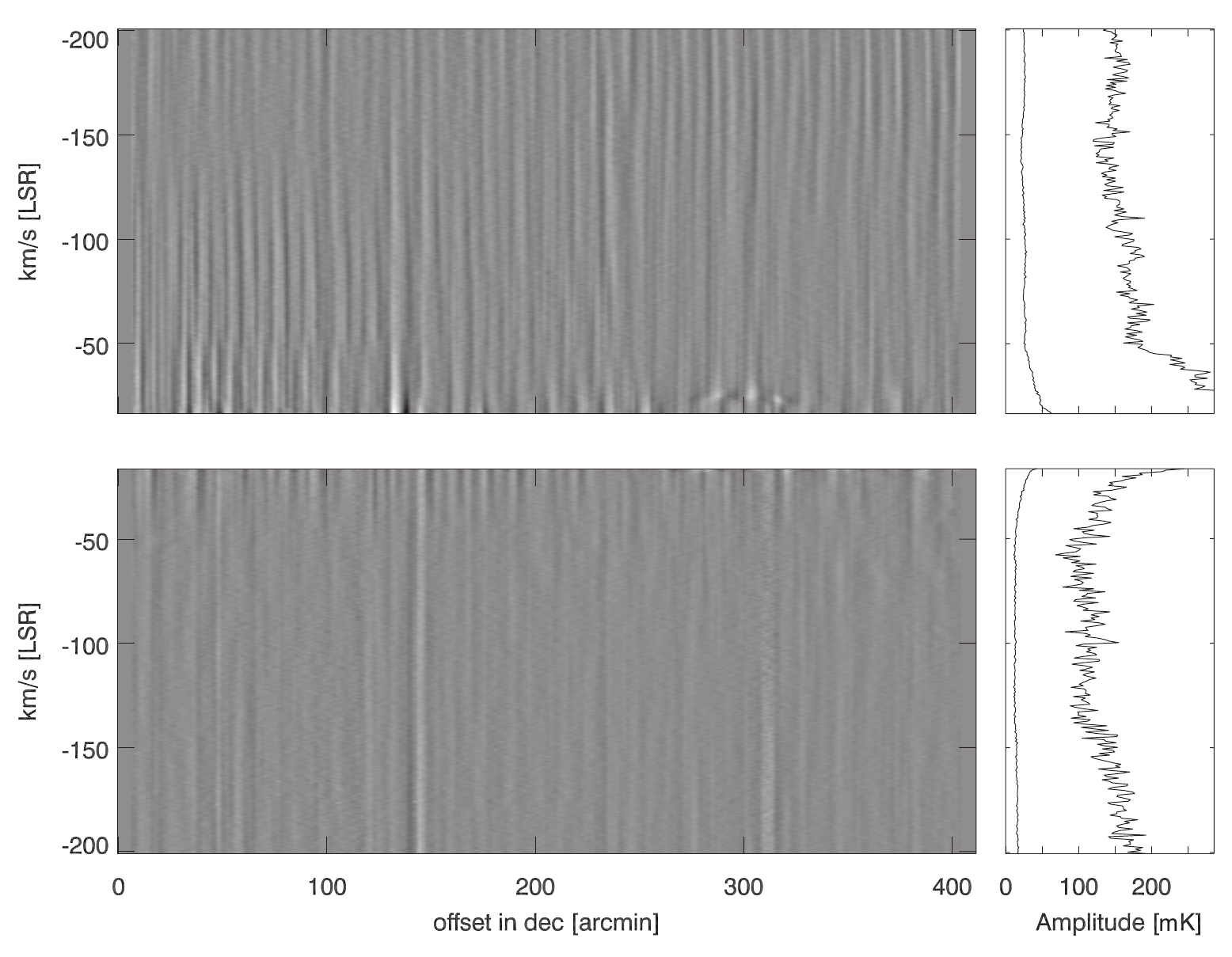}
\caption{Panels representing typical-to-strong residual FPN in the DR1-S (top) and DR1-F (bottom) data. The left panels are data cubes averaged over a few degrees in RA in high-latitude regions with typical strong FPN taken in drift mode. The data have had variation above a scale of 15$^\prime$ removed to highlight the beam-to-beam FPN variation. The right panels show the 1-$\sigma$ (lower, smooth) and peak-to-trough (higher, jagged) amplitude of the FPN as a function of velocity. The data show that DR1-S is somewhat worse than DR1-F, that the FPN errors increase somewhat as the velocity nears the \hi line, and that the typical FPN amplitude is 150 mK peak-to-trough, or 20 mK 1-$\sigma$ in these more contaminated regions. }
\label{fpndr1}
\end{center}
\end{figure}

Data reduction was very similar for both halves of DR1, but some differences exist. For DR1 a routine for removing radio intermodulation products was added for use with TOGS2 (commensal with GALFACTS; see Table \ref{dr1_proj}), because of the specific LO setup used. The exact protocol for data flagging was improved between DR1-S and DR1-F, which has led to improved data quality in DR1-F. Primarily residual fixed pattern noise is lower in DR1-F than in DR1-S, owing to improved algorithms and more precise by-eye checking procedures. An example of the scale of strong FPN residuals in the two halves of the data set can be seen in Figure \ref{fpndr1}. There are also fewer occasional RFI bursts and glitches in DR1-F, though these are rare in either data set. 

\begin{figure}
\begin{center}
\includegraphics[scale=0.95, angle=0]{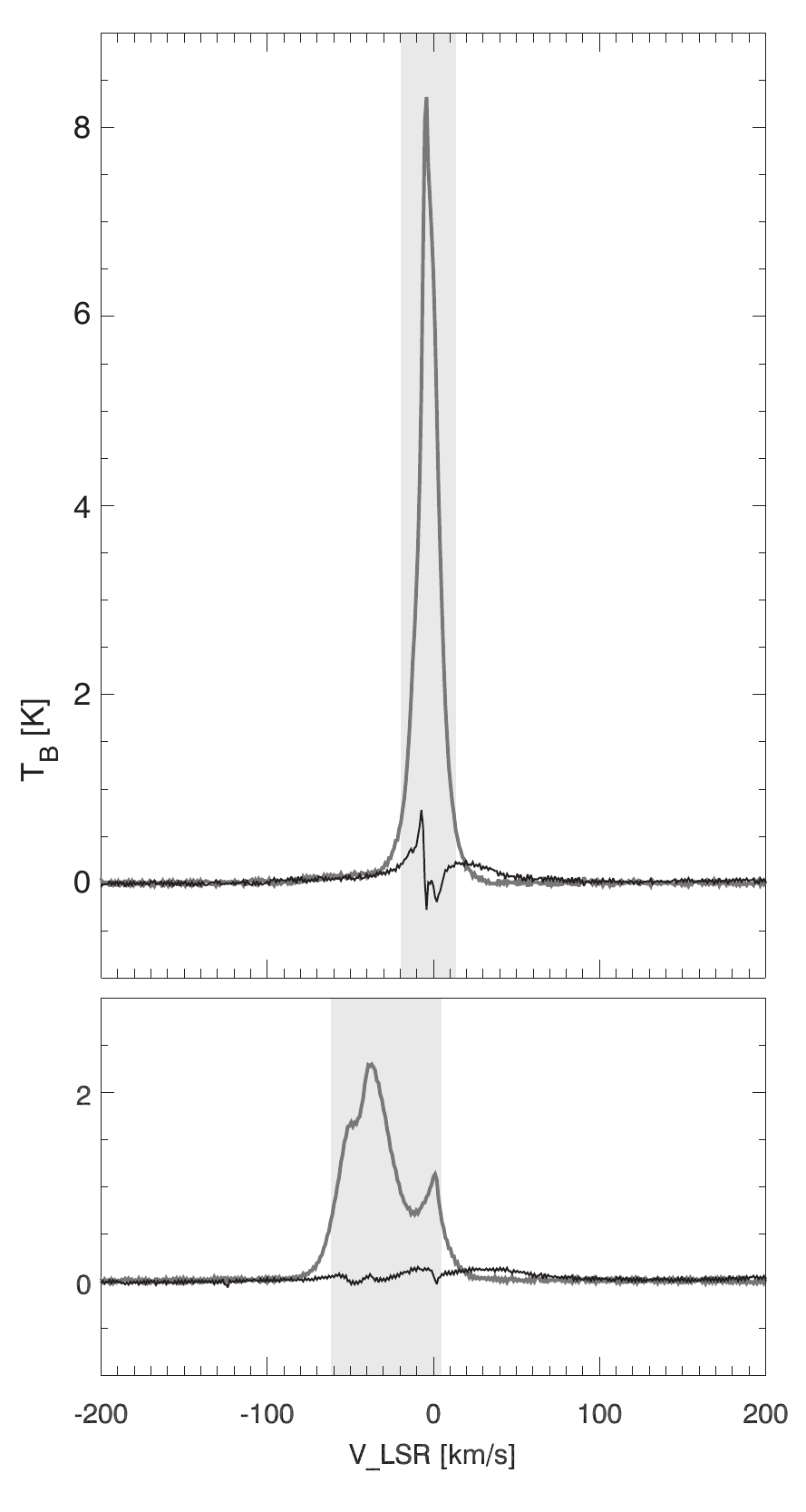}
\caption{Spectra toward $(l,b)=(210^{\circ},65^{\circ})$, a very low column region in DR1 (bottom) and $(l,b)=(45^{\circ},45^{\circ})$, a low column region observed while the bright Galactic plane is well above the horizon (top). The thick gray line is the LAB data, and the thinner black line is the GALFA-\hi data minus the LAB data, which outside regions of bright \hi represents the contamination by stray radiation towards these directions in the GALFA-\hi data. We have marked with a gray box the velocity range where we expect the difference to be dominated by beam mismatch between the LAB survey and our reconstructed LAB beam with GALFA-\hi data, and where we can therefore not use the plotted difference as an indicator of stray radiation strength.}\label{strayamp}
\end{center}
\end{figure}

The Arecibo 305 m is also known to have significant distant sidelobes, or stray radiation contamination. This effect can be important in Galactic \hi work, as the sky is bright at Galactic velocities in all directions. Since the stray sidelobes are large and low-gain, the effect is typically to add a wide, low amplitude signal that varies slowly with position. We intend to implement some stray radiation correction with future data releases, but for DR1 we simply attempt to parameterize the amplitude of the systematic error. Figure \ref{strayamp} shows spectra towards two positions in the sky, one of the lowest column parts of the DR1 sky ($l = 210^{\circ}$, $b = 65^{\circ}$), and one towards a low column part of the DR1 sky with the largest amount of bright Galactic \hi high above the horizon, to maximize stray contamination($l = 45^{\circ}$, $b = 45^{\circ}$). We show both the average LAB spectrum for the region and the difference between the GALFA-\hi average spectrum for the region and the LAB spectrum for the region to highlight the effect of stray radiation. We find that the stray effect is strongest closest to the line ($|V_{\rm LSR}| < 50$ \kmsa),  with an amplitude below 200 mK outside $\pm$ 20 \kms LSR, and that it is strongest when the Galactic plane is high in the sky. Within $\pm$ 20 \kms LSR, we cannot accurately enough match the beam shape and sensitivity profile of the LAB data to get a reliable measurement of the stray radiation effect, though it is unlikely to exceed 500 mK.

\section{Future Directions and Unresolved Issues}\label{futdir}
We have made significant strides towards reliable, artifact-free data, and we expect that these data will be a useful resource for the astronomy community. To further this goal we intend to continue to pursue known sources of systematic error in our data, including the effects of more distant sidelobes and stray radiation. 

\subsection{Further Sidelobe Calibration}
While the primary source of stray radiation is undoubtably the first sidelobes, the ALFA beam pattern is known to have significant power in secondary and higher-order sidelobes. These sidelobes roughly follow an Airy pattern model with elliptical cross section \citep{C-M2002}. Second sidelobes are more distant from the center of the observation, so while the contamination amplitude will be lower, the signal in the second sidelobe will be typically be more divergent from the main beam than in the first sidelobe. In cases where there is very fast variation in the sky signal this second sidelobe can have a significant effect, and we believe that it is currently the dominant contribution to uncorrected sidelobe radiation. 

To correct for the second and higher-order sidelobes these sidelobes must be carefully measured. Some raster maps have been made of the second sidelobe response, but they are significantly incomplete. To parameterize the basic components of second sidelobe we will measure it with the `spider scan' method outlined in \citet{Heiles04}. We will then add the second sidelobe response to our calibration method described in \S \ref{fsc}.

	\subsection{Stray Radiation Calibration}
We have shown that there is non-trivial stray radiation in our data. While the stray radiation does not pose a problem to conduct most of the science we are pursuing, including any science dependent upon small scale morphological features, or features at high velocity, it can be a problematic contaminant, especially when precise total \hi columns are needed. We plan to correct for this stray effect in future data releases by bootstrapping our results to the LAB data, and removing the differences we detect. This will allow us to treat the stray data properly without modeling the very complex feed response far from the beam center generated by the blocked Arecibo 305 m aperture. 

\section{Conclusion}

In this work we described the systems (\S \ref{thesys}), observing modes (\S \ref{obstech}) and data reduction methods (\S \ref{redproc}) used to produce the GALFA-\hi survey. We demonstrated the efficacy of these new and inventive techniques as well as their current limitations, and showed the high quality of the resulting data products (\S \ref{exr}). We presented the GALFA-\hi Data Release 1 (\S \ref{dr1}) and discussed methods for improving our final data product in the future (\S \ref{futdir}). We found that we were capable of dramatically reducing the impact of these systematic effects, in many cases making them undetectable.  

The DR1 data product can be found at https://purcell.ssl.berkeley.edu/, downloadable either in the original survey data cubes or in any user-defined region and spectral resolution. The original data cubes are 512$^\prime$ in small-circle Right Ascension and Declination, with 2048 channels, either at the original resolution of 0.184 \kms with a velocity range from $-188$ \kms to $188$ \kms LSR or at a degraded resolution of 0.736 \kms over the full bandwidth.

GALFA-\hi is very much an ongoing survey with hours of data still being taken daily through commensal and dedicated projects. As such, future data releases are planned including these new data and improved reduction methodologies, and data release 2 is tentatively planned for late 2011. Many astronomers around the world are engaged in active research with GALFA-\hi data, and we plan to continue to provide high quality data to these astronomers as well as the rest of the astronomical community in the future. 
	
\section{Acknowledgements}

The authors would like to thank the many, many people who made this work, and the ongoing survey it represents, possible. We thank Paul Goldsmith, Tom Bania and John Dickey for early science conceptual work on the GALFA project, and the entire GALFA-\hi collaboration for significant input over the lifetime of the project. We thank the incredibly helpful and dogged Arecibo staff, including the telescope operators who worked with us at great length to develop our observing protocols. In particular we would like to thank Phil Perillat and Mikael Lerner for their contributions to the GALFA-\hi data reduction pipeline and observing codes, respectively. We thank Arun Venkataraman for assistance in data handling and storage. We would also like to thank Dan Wertheimer and his team for designing and building the GALFA spectrometer, GALSPECT. We would like to especially thank the late Jeff Mock, the driving force behind GALSPECT, without whose dedication and brilliance this project could not have happened. This work was supported in part by NSF grant AST-0406987 and NSF Collaborative Research grants AST-0707679,070758,0709347,0917810, as well as Research Corporation and Herschel grant 1369759. KAD acknowledges funding from the European Community's Seventh Framework Program under grant agreement n$^{\rm o}$ PIIF-GA-2008-221289.

\bibliographystyle{apj}


\end{document}